\input harvmac.tex

\input epsf.tex

\def\figin{\epsfcheck\figin}\def\figins{\epsfcheck\figins}
\def\epsfcheck{\ifx\epsfbox\UnDeFiNeD
\message{(NO epsf.tex, FIGURES WILL BE IGNORED)}
\gdef\figin##1{\vskip2in}\gdef\figins##1{\hskip.5in}
\else\message{(FIGURES WILL BE INCLUDED)}%
\gdef\figin##1{##1}\gdef\figins##1{##1}\fi}
\def\DefWarn#1{}
\def\figinsert{\goodbreak\midinsert}
\def\ifig#1#2#3{\DefWarn#1\xdef#1{fig.~\the\figno}
\writedef{#1\leftbracket fig.\noexpand~\the\figno}%
\figinsert\figin{\centerline{#3}}\medskip\centerline{\vbox{\baselineskip12pt
\advance\hsize by -1truein\noindent\footnotefont{\bf
Fig.~\the\figno:} #2}}
\bigskip\endinsert\global\advance\figno by1}



\lref\firstconf{
  J.~M.~Drummond, J.~Henn, V.~A.~Smirnov and E.~Sokatchev,
  ``Magic identities for conformal four-point integrals,''
  JHEP {\bf 0701}, 064 (2007)
  [arXiv:hep-th/0607160].
}

 \lref\karch{
 A.~Karch and E.~Katz,
  ``Adding flavor to AdS/CFT,''
  JHEP {\bf 0206}, 043 (2002)
  [arXiv:hep-th/0205236].
  }
\lref\pineda{
  A.~Pineda,
  ``The static potential in {\cal N}=4 supersymmetric Yang-Mills at weak
  coupling,''
  arXiv:0709.2876 [hep-th].
}
\lref\dineetal{
  T.~Appelquist, M.~Dine and I.~J.~Muzinich,
  ``The Static Limit Of Quantum Chromodynamics,''
  Phys.\ Rev.\  D {\bf 17}, 2074 (1978).
}
\lref\grossetal{
  N.~Drukker, D.~J.~Gross and H.~Ooguri,
  ``Wilson loops and minimal surfaces,''
  Phys.\ Rev.\  D {\bf 60}, 125006 (1999)
  [arXiv:hep-th/9904191].
}

\lref\zarembowillo{
  J.~K.~Erickson, G.~W.~Semenoff and K.~Zarembo,
  ``Wilson loops in N = 4 supersymmetric Yang-Mills theory,''
  Nucl.\ Phys.\  B {\bf 582}, 155 (2000)
  [arXiv:hep-th/0003055].
}
\lref\bds{
  Z.~Bern, L.~J.~Dixon and V.~A.~Smirnov,
  ``Iteration of planar amplitudes in maximally supersymmetric Yang-Mills
  theory at three loops and beyond,''
  Phys.\ Rev.\  D {\bf 72}, 085001 (2005)
  [arXiv:hep-th/0505205].
}

\lref\bes{ N.~Beisert, B.~Eden and M.~Staudacher,
  ``Transcendentality and crossing,''
  J.\ Stat.\ Mech.\  {\bf 0701} (2007) P021
  [arXiv:hep-th/0610251].
}

\lref\fajmtwo{  L.~F.~Alday and J.~Maldacena,
  ``Comments on operators with large spin,''
  arXiv:0708.0672 [hep-th].
}

\lref\fajm{
  L.~F.~Alday and J.~Maldacena,
  ``Gluon scattering amplitudes at strong coupling,''
  JHEP {\bf 0706}, 064 (2007)
  [arXiv:0705.0303 [hep-th]].
}

\lref\drukkerwil{
  N.~Drukker and S.~Kawamoto,
  ``Small deformations of supersymmetric Wilson loops and open spin-chains,''
  JHEP {\bf 0607}, 024 (2006)
  [arXiv:hep-th/0604124].
  }

\lref\gkpfirst{
  S.~S.~Gubser, I.~R.~Klebanov and A.~M.~Polyakov,
  ``Gauge theory correlators from non-critical string theory,''
  Phys.\ Lett.\  B {\bf 428}, 105 (1998)
  [arXiv:hep-th/9802109].
}

\lref\ansar{  O.~Aharony, A.~Fayyazuddin and J.~M.~Maldacena,
  ``The large N limit of N = 2,1 field theories from three-branes in
  F-theory,''
  JHEP {\bf 9807}, 013 (1998)
  [arXiv:hep-th/9806159].
  }

\lref\brandhuber{ A.~Brandhuber, P.~Heslop and G.~Travaglini,
  ``MHV Amplitudes in N=4 Super Yang-Mills and Wilson Loops,''
  arXiv:0707.1153 [hep-th].
 }
\lref\kruczenski{ M.~Kruczenski,
  ``A note on twist two operators in N = 4 SYM and Wilson loops in Minkowski
  signature,''
  JHEP {\bf 0212}, 024 (2002)
  [arXiv:hep-th/0210115].
  }
\lref\gkp{  S.~S.~Gubser, I.~R.~Klebanov and A.~M.~Polyakov,
  ``A semi-classical limit of the gauge/string correspondence,''
  Nucl.\ Phys.\  B {\bf 636}, 99 (2002)
  [arXiv:hep-th/0204051].
}

\lref\otherfun{
 Z.~Bern, M.~Czakon, L.~J.~Dixon, D.~A.~Kosower and V.~A.~Smirnov,
  Phys.\ Rev.\  D {\bf 75}, 085010 (2007)
  [arXiv:hep-th/0610248].
 S.~Frolov and A.~A.~Tseytlin,
  JHEP {\bf 0206}, 007 (2002)
  [arXiv:hep-th/0204226].
 R.~Roiban and A.~A.~Tseytlin,
  arXiv:0709.0681 [hep-th].
   M.~K.~Benna, S.~Benvenuti, I.~R.~Klebanov and A.~Scardicchio,
  Phys.\ Rev.\ Lett.\  {\bf 98}, 131603 (2007)
  [arXiv:hep-th/0611135].
  B.~Basso, G.~P.~Korchemsky and J.~Kotanski,
  arXiv:0708.3933 [hep-th].
  }

\lref\morefl{
 S.~Abel, S.~Forste and V.~V.~Khoze,
  arXiv:0705.2113 [hep-th].
 E.~I.~Buchbinder,
  arXiv:0706.2015 [hep-th].
 M.~Kruczenski, R.~Roiban, A.~Tirziu and A.~A.~Tseytlin,
  arXiv:0707.4254 [hep-th].
    A.~Jevicki, C.~Kalousios, M.~Spradlin and A.~Volovich,
  arXiv:0708.0818 [hep-th].
  Z.~Komargodski and S.~S.~Razamat,
  ``Planar quark scattering at strong coupling and universality,''
  arXiv:0707.4367 [hep-th].
 A.~Mironov, A.~Morozov and T.~N.~Tomaras,
  arXiv:0708.1625 [hep-th].
  H.~Kawai and T.~Suyama,
  arXiv:0708.2463 [hep-th].
  S.~G.~Naculich and H.~J.~Schnitzer,
  arXiv:0708.3069 [hep-th].
  }

\lref\mcgreevy{
 J.~McGreevy and A.~Sever,
  ``Quark scattering amplitudes at strong coupling,''
  arXiv:0710.0393 [hep-th].
}

\lref\jmwilson{
  J.~M.~Maldacena,
  ``Wilson loops in large N field theories,''
  Phys.\ Rev.\ Lett.\  {\bf 80}, 4859 (1998)
  [arXiv:hep-th/9803002]. 
 }
\lref\wilsonrey{
  S.~J.~Rey and J.~T.~Yee,
  ``Macroscopic strings as heavy quarks in large N gauge theory and  anti-de
  Sitter supergravity,''
  Eur.\ Phys.\ J.\  C {\bf 22}, 379 (2001)
  [arXiv:hep-th/9803001].
}

\lref\nfive{
  J.~M.~Drummond, J.~Henn, G.~P.~Korchemsky and E.~Sokatchev,
  ``On planar gluon amplitudes/Wilson loops duality,''
  arXiv:0709.2368 [hep-th].
}
\lref\zarembo{
  J.~K.~Erickson, G.~W.~Semenoff, R.~J.~Szabo and K.~Zarembo,
  ``Static potential in N = 4 supersymmetric Yang-Mills theory,''
  Phys.\ Rev.\  D {\bf 61}, 105006 (2000)
  [arXiv:hep-th/9911088].
}
\lref\wilsonplanar{
  J.~M.~Drummond, G.~P.~Korchemsky and E.~Sokatchev,
  ``Conformal properties of four-gluon planar amplitudes and Wilson loops,''
  arXiv:0707.0243 [hep-th].
}

\lref\DrummondBY{
  J.~M.~Drummond, L.~Gallot and E.~Sokatchev,
  ``Superconformal invariants or how to relate four-point AdS amplitudes,''
  Phys.\ Lett.\  B {\bf 645}, 95 (2007)
  [arXiv:hep-th/0610280].
}

\lref\DrummondRZ{
  J.~M.~Drummond, J.~Henn, V.~A.~Smirnov and E.~Sokatchev,
  ``Magic identities for conformal four-point integrals,''
  JHEP {\bf 0701}, 064 (2007)
  [arXiv:hep-th/0607160].
}

\lref\vanishingamp{
  M.~T.~Grisaru and H.~N.~Pendleton,
  ``Some Properties Of Scattering Amplitudes In Supersymmetric Theories,''
  Nucl.\ Phys.\  B {\bf 124}, 81 (1977).
 S.~J.~Parke and T.~R.~Taylor,
  ``Perturbative QCD Utilizing Extended Supersymmetry,''
  Phys.\ Lett.\  B {\bf 157}, 81 (1985)
  [Erratum-ibid.\  {\bf 174B}, 465 (1986)].
}

\lref\nonabelian{
  J.~G.~M.~Gatheral,
  ``Exponentiation Of Eikonal Cross-Sections In Nonabelian Gauge Theories,''
  Phys.\ Lett.\  B {\bf 133}, 90 (1983).
}

\lref\NguyenYA{
  D.~Nguyen, M.~B.~Spradlin and A.~Volovich,
  ``New Dual Conformally Invariant Off-Shell Integrals,''
  arXiv:0709.4665 [hep-th].
}

\lref\GubserTV{
  S.~S.~Gubser, I.~R.~Klebanov and A.~M.~Polyakov,
  ``A semi-classical limit of the gauge/string correspondence,''
  Nucl.\ Phys.\  B {\bf 636}, 99 (2002)
  [arXiv:hep-th/0204051].
}

\lref\KotikovER{
  A.~V.~Kotikov, L.~N.~Lipatov, A.~I.~Onishchenko and V.~N.~Velizhanin,
  ``Three-loop universal anomalous dimension of the Wilson operators in N =  4
  SUSY Yang-Mills model,''
  Phys.\ Lett.\  B {\bf 595}, 521 (2004)
  [Erratum-ibid.\  B {\bf 632}, 754 (2006)]
  [arXiv:hep-th/0404092].
}

\lref\DrukkerEP{
  N.~Drukker, D.~J.~Gross and A.~A.~Tseytlin,
  ``Green-Schwarz string in AdS(5) x S(5): Semiclassical partition  function,''
  JHEP {\bf 0004}, 021 (2000)
  [arXiv:hep-th/0001204].
}

\lref\furtherconformal{
  Z.~Bern, M.~Czakon, L.~J.~Dixon, D.~A.~Kosower and V.~A.~Smirnov,
  ``The Four-Loop Planar Amplitude and Cusp Anomalous Dimension in Maximally
  Supersymmetric Yang-Mills Theory,''
  Phys.\ Rev.\  D {\bf 75}, 085010 (2007)
  [arXiv:hep-th/0610248].
  Z.~Bern, J.~J.~M.~Carrasco, H.~Johansson and D.~A.~Kosower,
  ``Maximally supersymmetric planar Yang-Mills amplitudes at five loops,''
  arXiv:0705.1864 [hep-th].
}

\Title{\vbox{\baselineskip12pt \hbox{SPIN-07/41} \hbox{
ITP-UU-07/55} }} {\vbox{\centerline{ Comments  on gluon scattering
amplitudes via AdS/CFT }}}
\bigskip
\centerline{ Luis F. Alday$^{a}$ and Juan Maldacena$^b$}
\bigskip
\centerline{\it $^a$Institute for Theoretical Physics and Spinoza
Institute} \centerline{Utrecht University, 3508 TD Utrecht, The
Netherlands}

\centerline{ \it  $^b$School of Natural Sciences, Institute for
Advanced Study} \centerline{\it Princeton, NJ 08540, USA}

\vskip .3in \noindent
In this article we consider $n$ gluon color ordered, planar amplitudes in
${\cal N}=4$ super Yang Mills at strong 't Hooft coupling. These amplitudes
are approximated by classical surfaces in $AdS_5$ space. We compute the value
of the amplitude for a particular kinematic configuration for a large number of
gluons and
find that the result disagrees with a recent guess for the exact value of the amplitude.
Our results are still compatible with a possible relation between amplitudes and Wilson loops.

In addition, we also give a prescription for computing processes
involving local operators and asymptotic states with a fixed
number of gluons. As a byproduct, we also obtain a string theory prescription for computing
the dual of the ordinary Wilson loop, $Tr P e^{ i\oint A } $, with no couplings to the
scalars. We also evaluate the quark-antiquark potential at two loops.


 \Date{ }


\newsec{Introduction}

In this paper we investigate some properties  of planar gluon
scattering amplitudes in ${\cal N}=4$ super
Yang Mills at strong coupling.
 We use the prescription in \fajm\ (see also \refs{\morefl,\mcgreevy}) for computing $n$
gluon scattering amplitudes at strong coupling. The computation of
the amplitude reduces to a geometric problem. One has to find a
surface in $AdS_5$ which ends on the boundary on a sequence of
light-line lines specified by the momenta of the gluons $\{ k_1,
k_2 , \cdots, k_n \}$. In \fajm , we computed the explicit form of
the amplitude for the scattering of four gluons, which matched the
conjectured form in \bds . At strong coupling the problem has a
``dual" conformal symmetry which is enough for determining the
four and five gluon scattering amplitudes \refs{\fajm,\nfive}.
This symmetry is also present in weak coupling computations
\refs{\firstconformal,\wilsonplanar,\nfive,\NguyenYA}, but its
origin and full scope remain mysterious\foot{In other words,
despite the great deal of evidence obtained by direct computation
of the amplitudes \refs{\firstconformal,\furtherconformal}, we do
not know whether this is an exact symmetry of the planar theory.}.

One of the goals of this article is to perform a computation which
is not determined by the ``dual space" conformal symmetry. This
only starts happening when one considers $n\geq 6$ gluons \nfive ,
where one can start writing non-vanishing conformal invariant
cross ratios of the momenta. Doing explicit computations in this
case seems rather difficult, because it is hard to find the
minimal area surfaces. However, here we will consider the case of
$n\to \infty$ where the problem simplifies. The reason is that we
can consider a configuration of lightlike segments which
approximates a spacelike line. For some spacelike configurations
we can find the minimal surfaces directly, as in
\refs{\jmwilson,\wilsonrey}.
 We use this result to test the BDS conjecture
\bds\ which, as shown in  \refs{\wilsonplanar,\brandhuber},
amounts to exponentiating the one loop result for a Wilson loop
expectation value, up to an overall coefficient in the exponent
which is given by the cusp anomalous dimension. Using the relation
between the weak and strong coupling forms of the quark-anti-quark
potential,
 one can check that the BDS conjecture is not
correct at strong coupling and for a large number of gluons.
Recently, a separate conjecture has been discussed \refs{\wilsonplanar,\brandhuber,\nfive},
 which states that  the finite part of MHV amplitudes
is the same as  the finite part of Wilson loop expectation values. Our results do not
offer new arguments for or against this relationship.

As a byproduct of the approximation of spacelike Wilson loop by a zig-zag sequence
of light-like ones we get a prescription for the strong coupling dual of the Wilson
loop operator with no couplings to the scalar. It is basically the same prescription
as in \refs{\jmwilson,\wilsonrey} except that we have Neumann boundary conditions on
the $S^5$.

In this paper we also consider the problem of computing amplitudes involving the
insertion of an operator and an $n$ gluon final state. Such processes arise
when we excite the theory and produce final gluon states.
We consider the description of these exclusive processes at strong coupling and
we will give a prescription for computing them.

This paper is organized as follows. In section two we  describe
the strong coupling computation of a process involving and
operator insertion ${\cal O}$ and an $n$ gluon final state. We start with
this discussion because the surfaces that appear here will be relevant later, but the
casual reader can skip this section.
In section three we consider the $n$ gluon scattering amplitude for large $n$ and a
very specific kinematic configuration which is chosen so that it approximates a
simple spacelike Wilson loop. We use this configuration to test the BDS ansatz \bds\
at strong coupling and we find a disagreement.
In section four we discuss some aspects of the Wilson loop operator with no
coupling to the scalars and we perform some weak and strong coupling computations for this
Wilson loop.
We finally present some conclusions. In appendix A we discuss some aspects of the five
gluon solution which we did not find explicitly. In appendix B we give some details for
the perturbative computations we did for Wilson loops.

\newsec{ Processes involving a local operator and asymptotic gluons states }

\subsec{Short review on gluon scattering amplitudes at strong coupling}

\ifig\segments{Sequence of light-like segments on
the $AdS_5$ boundary given by the set of momenta.
 The string surface should end on this line at the boundary of $AdS_5$.
In this case we have   six gluons.} {\epsfxsize1.5in\epsfbox{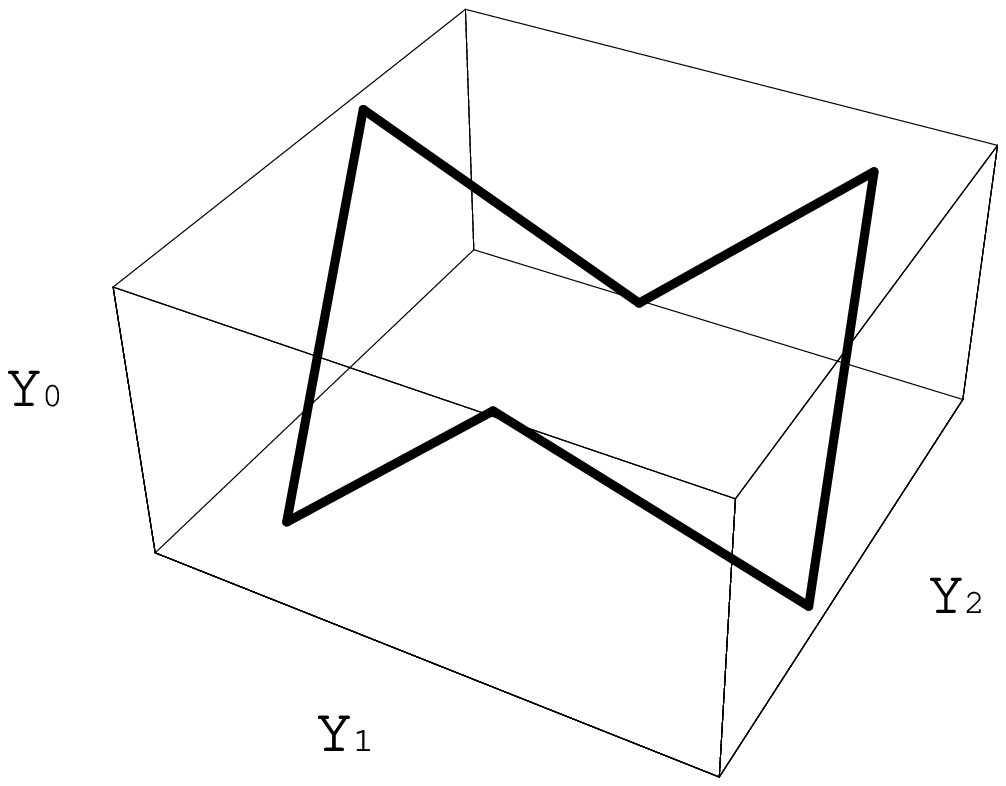}}

Planar color ordered gluon scattering amplitudes can be computed
at strong coupling by considering classical string  in $AdS_5$ \fajm .
The amplitude is given in terms of the area of a minimal surface in $AdS_5$
 \eqn\finas{
 {\cal A} \sim  e^{ - { \sqrt{ \lambda } \over 2 \pi } ({\rm Area} ) }
} where  we set the  radius of curvature of $AdS_5$ to one. This
is a minimal surface which ends at the boundary of $AdS_5$ space
on a sequence of lightlike segments given by the four-momenta of
the gluons, see \segments . The information
 about the polarization states of the gluons (or other states
of the ${\cal N}=4$ super Yang Mills multiplet) comes in at
 higher orders in the $1/\sqrt{\lambda}$
expansion. In fact, the next correction to \finas\  is  a
$\lambda$ independent factor multiplying \finas .  Such a factor
could vanish for specific external states. For instance, tree
level amplitudes with all helicities identical or all but one
identical, vanish due to supersymmetry Ward identities
\vanishingamp\ .

The $AdS_5$ space that appears in the above prescription is parametrized by coordinates
$y_\mu$, $\mu =0,1,2,3$ and $r$ with the metric
\eqn\metr{
ds^2 = { dy^2 + dr^2 \over r^2 }
}
This space is related to the ordinary $AdS_5$ space appearing as the gravity dual
of ${\cal N}=4 $ super Yang Mills by a T-duality along the four  worldvolume directions
 and
by $r=1/z$ where
\eqn\adsfiveme{
ds^2 = { dx^2 + dz^2 \over z^2 }
}
is the metric of the ordinary $AdS_5$ geometry describing the gravity dual of
 ${\cal N}=4$ super Yang Mills and $x^\mu$ are the coordinates of the $R^{1,3}$ space
 where the field theory lives.
More details  can be found in \fajm .

\subsec{Processes involving asymptotic  gluons and local operators }

In this section we would like to give a prescription for computing
amplitudes for processes involving asymptotic gluon states and local operators.
 For example we can  excite the theory by a local
 operator and  look at final states containing a certain   number of  gluons.
 We have in mind processes similar to the ones that arise when we
 consider $e^+ e^- \to \gamma \to $jets. In this case, we can
 analyze the process to lowest order in $\alpha_{em}$ but to all
 orders in $\alpha_{strong}$ by noticing that the photon couples
 to the electromagnetic current of QCD and this in turn can
 produce various final states. Thus the final
 hadronic state is produced by acting with the QCD electromagnetic current
on the vacuum. To lowest order in
 $\alpha_{strong}$ the   state is, of course, a quark
 anti-quark pair.

 We now want to consider analogous processes in ${\cal N}=4$ super
 Yang Mills at strong coupling in the planar approximation. Thus we consider
 a process where
 we add a local  operator to the theory and we produce gluons. The
 local operator is a single trace operator with given momentum
 \eqn\locaop{
 {\cal O}(q) = \int d^4x e^{ i q.x} {\cal O}(x)
 }
 We can consider any operator of the theory. Concrete examples are
 the stress tensor, the R-symmetry currets, etc.

We are interested in exclusive final states consisting of
individual gluons, or other members of the ${\cal N}=4$
supermultiplet. From now on the word ``gluon'' will mean any
element of the supermultiplet: a gluon, fermion or scalar, all in the adjoint
representation.
The asymptotic states for these colored objects
 are
 only well defined after we use an IR regulator. The simplest one
 is dimensional regularization, which consists in going to
 $4+\epsilon$ dimensions. Then the theory is free in the IR and
 gluons  are good
 asymptotic states. On the gravity side this can be done by
 considering the near horizon metrics of D-$p$ branes with
 $p=3+\epsilon$, as explained in \fajm .

 Once we regularize, we   have a worldsheet whose
boundary conditions in the far past or future are set at $z\sim
\infty$, where the asymptotic gluons live, and the operator
conditions are set at the boundary of $AdS_5$,  $z \sim 0$, in \adsfiveme .
In the T-dual  coordinates \metr\ the
asymptotic  states carry winding number which is proportional to the
momentum. The gluon final states are represented as in \fajm\ by
considering a sequence of lightlike lines at $r=0$. Each
light-like segment joints two points separated by  $ 2 \pi
k^\mu_i$. As opposed to the situation considered in \fajm , this
sequence is not closed. In fact we have $\sum_{i=1}^n k_i^\mu = q^\mu$
where $q^\mu$ is the momentum of the operator. It is convenient to
formally think of the coordinate along $q^\mu$ as compact and to consider
a  closed string as winding that coordinate. This is
equivalent to saying that we consider an infinite periodic
superposition of the set of momenta $\{ k_1,k_2, \cdots , k_n \}$.

We now should give a prescription for the operator. An operator
insertion  leads to a string that goes to the $AdS_5$ boundary,
$z=0$ in the coordinates \adsfiveme.
 This implies that it should
go to $r=\infty$ in the dual coordinates \metr .
 Thus we consider a string stretching along
the direction $q^\mu$ that goes to $r=\infty$.

As a simple example, consider a two gluon
 state and an operator insertion.
 The two gluon momenta obey  $k^\mu_1 + k^\mu_2 = q^\mu$.
Let us consider the case where the mometum $q^\mu$ is
spacelike and $k_1$ is incoming and $k_2$ outgoing. By performing a boost
we can
 choose the momenta as
\eqn\momenta{ (2 \pi )k_1^\mu = ({\kappa\over 2},{\kappa \over
2},0) ~,~~~~~~~(2 \pi ) k_2^\mu = (-{\kappa \over 2},{\kappa \over
2}, 0) ~,~~~~~~(2 \pi ) q^\mu  =(0,\kappa,0) }
 It is convenient to view the direction
$y^1$ as a compact direction with period $y^1  \sim y^1 +  \kappa$
so that the total winding number of the string corresponds to an
allowed closed string. This closed string has to end on the
boundary of the original $AdS_5$ space \adsfiveme\ at $z=0$. In
terms of the dual metric \metr\ it should go to $r =\infty$.
\ifig\zigzag{(a) We consider the configuration of light-like lines
corresponding to the initial and final state gluons under
consideration. In (b) we consider an infinite repetition of the
configuration. In (c) a more general configuration with four
gluons is considered. And in (d) we draw the corresponding
periodic version.
 } {\epsfxsize2.5in\epsfbox{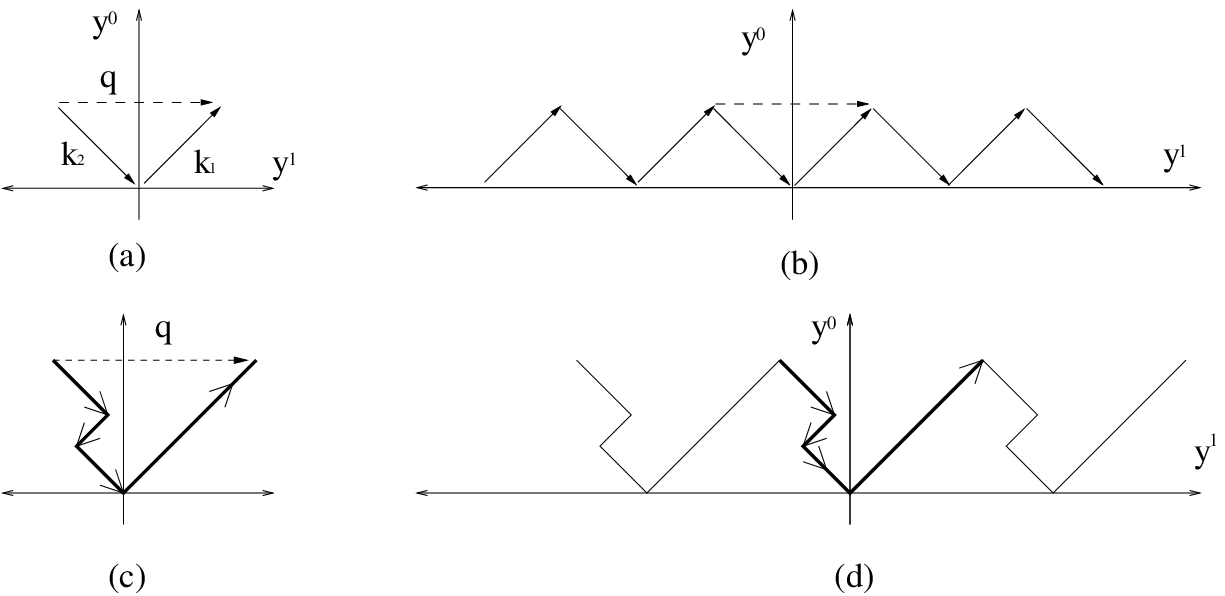}}

In order to find the surface it is convenient
   to  consider an infinite repetition of these momenta, which
are following a zig-zag path in the $y^0,y^1$ plane as shown in
figure 2. We have discussed the solution by thinking of the
direction $y_1$ as compact. By the time that we undo the T-duality
and we go back to the original $AdS_5$ coordinates $x,z$ we can
forget about the fact that the coordinate is compact. In other
words, the final solution, back in $x,z$ coordinates,
 also describes the case where the $x$
coordinates are  non-compact.

We  look for a worldsheet which is extended in the radial $AdS_5$
direction, from $r=0$, where it ends on the  contour displayed in
\zigzag (b), and extends all the way to $r\to \infty$. As we go to
large $r$ the surface is extended in the $\hat 1$ spatial
direction but is localized in time. See figure 3 for a picture of
the expected surface.

\ifig\approxzigzag{Approximate form of the solution. At $r=0$ the surface ends on a zig-zag, while for large $r$, $t$ decays exponentially.
 } {\epsfxsize2.5in\epsfbox{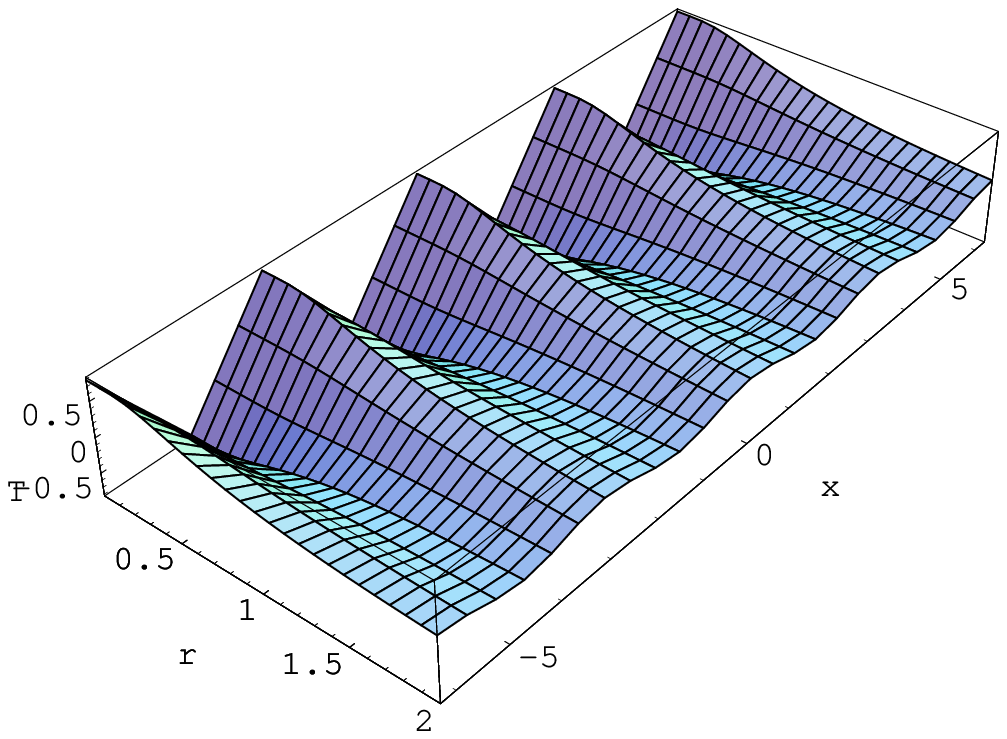}}

The amplitude is  given by computing the area over one period
of the resulting surface.  Let us point out some features of the solution,
which we have not found explicitly.
  First, one can write the Nambu-Goto action
by choosing $r,y$ as worldsheet coordinates so that $t(r,y)$ is
the unknown function. The action is \eqn\actng{
 iS = - { R^2 \over 2 \pi \alpha'} \int dy dr {
 \sqrt{ 1 - (\partial_y t)^2 - (\partial_r t)^2}
 \over r^2 }
}
Thus, we should solve the equations of motion that come from this action with the
boundary conditions
\eqn\boundcon{\eqalign{
 & t|_{r=\infty} =0,
 \cr
 & t|_{r=0} = y ~,~~~{\rm for} ~~~|y| \leq  {\kappa \over 4} ~,~~~~~~~~t_{r=0} =
 {\kappa \over 2} -y ~,~~~~~
 {\rm for}~~~ { \kappa \over 4 }  \leq y \leq  { 3\kappa \over 4}
 }}
 and extended in a periodic way outside this range, $t(r,y + \kappa )=t(r,y)$. The second line
 in \boundcon\ is simply specifying the path shown in \zigzag (b).

 This equation could, in principle, be solved numerically.
  We expect
 that for distances much bigger than the size of the zig-zag, {\it
 i.e.} $r \gg \kappa $, $t(r,y)$ is very small and satisfies a linear
 equation obtained by expanding \actng\ for small $t$.
 Expanding $t$ in Fourier modes, $t(r,y)=\sum_n
 t_{n}(r)e^{i k_n y}$, with $k_n=2 \pi n/\kappa$, we obtain the following equation for
 $t_{n}(r)$
 \eqn\tsmall{
  -k_n^2 t_{n}(r)+r^2 \partial_r \left[ {1 \over r^2} \partial_r t_{n}(r)\right]=0~~~~\rightarrow~~~~
  t_{n}(r)= c_{n}  e^{-k_n r}(k_n r+1)
  }
where we have kept only the decaying solution (for positive
$k_n$). Note that due to the exponential decay, already when $r$
is a few times bigger than $|\kappa|$, the above solution will be
a good approximation. The coefficients $c_{n}$ are determined by
imposing the boundary condition at $r=0$ \boundcon , but we should
recall that we cannot use the linearized equation in that region.
The problem has a scaling symmetry that implies that we can scale
out $\kappa$ so that the solution is \eqn\periodict{
t(r,y)=\kappa~ \hat{t}({r \over \kappa},{y \over \kappa}) \equiv
\kappa ~\hat{t}(\hat{r},\hat{y}) } It is then easy to see, that
the value of the classical action \actng\  on this solution is
formally independent of the scale $\kappa$, as expected from
scaling symmetry. However, since there is a divergence, an
explicit $\kappa$ dependence is introduced when we subtract  the
divergence. Let us understand the divergencies. Let us first
consider the large $r$ region.
 The integral in the region of large $r$ converges since $t\to 0$ so that
 we are simply  integrating
 $dr/r^2$. This might seem a bit surprising since we expected to obtain terms of the
 form $\log r$ that are related to the anomalous dimension of the operator.
 Notice, however, that a logarithmic term in the classical area would have implied
 an anomalous dimension of order $\sqrt{\lambda}$. Thus, the boundary conditions we
 considered correspond to operators whose anomalous dimension vanishes at this order.
 For an operator such as the stress tensor,
 which has anomalous dimension equal to four, this
 is indeed the case. We expect to obtain logarithmic terms when we go to higher order in
 the $1/\sqrt{\lambda}$ expansion.\foot{ An example of  a configuration where
 we get an anomalous dimension at leading order is the following.
  Consider a string
 winding on the sphere (for example we can replace
 $S^5 \to S^5/Z_k$ where the $Z_k$ acts without fixed points).
 In that
 case we will have the action $ {\sqrt{\lambda} \over 2 \pi }
 \int { dy dr \over r^2} \sqrt{ 1 + r^2 (\partial_y \theta)^2 }$.
 Expanding the square root for $r\to \infty$ we get ${\sqrt{\lambda} \over 2 \pi}
 \int dy {dr \over r} |\partial_y \theta| \sim {\sqrt{\lambda} \over 2 \pi}
 ( \Delta \theta ) \log r$. This corresponds to an operator of dimension $\Delta = { \sqrt{ \lambda} \over
 2 \pi } \Delta \theta $ which is indeed what we obtain for a string stretching on the sphere over an
 angle $\Delta \theta$. }

We can now consider the small $r$ region. The analysis of this
region is the same as the analysis in the small $r$ region for the
gluon scattering amplitudes discussed in \fajm . One can
dimensionally regularize the problem by going to $d = 4 - 2
\epsilon$ dimensions. Then the lagrangian becomes \eqn\newlag{ L =
{ \sqrt{ \lambda} \over 2 \pi} c_\epsilon \kappa^{-\epsilon} \int
d \hat r d \hat y {\hat  r}^{-\epsilon} {\cal L}_0[\hat t(\hat y,
\hat r)] } where we have rescaled all variables so that the only
dependence on $\kappa$ is in the overall factor. In \newlag\
$c_\epsilon$ is a function of only $\epsilon$.
 The divergencies
 arise from the region near the cusps connecting the momenta
of two adjacent gluons (see \bds ) and they can be computed using
the single cusp solution considered in \refs{\kruczenski,\fajm}.
The value of the action is given by integrating only over one
period in $y$. It evaluates to a function of the form \eqn\funfom{
i S = -{ \sqrt{ \lambda} \over 2 \pi}  { \mu^\epsilon \over ( 2
\pi \kappa) ^{\epsilon} }  \left[ 2 \left({ 1  \over \epsilon^2 }
+ { 1- \log 2   \over 2  \epsilon} \right)  + C \right] } The
coefficients of the divergent terms are locally determined and are
the same as in \fajm , so that we would only need the solution to
compute the constant $C$. For the simplest case of two gluons, the
solution does not depend on any kinematical variable. As we
consider configurations with more gluons the solution, and the
value of the amplitude, will start depending on the kinematic
invariants.

\subsec{Processes involving a mesonic operator and final quark and
anti-quarks}

In this subsection we consider a small variant of the configuration considered above.
We consider a large $N$ theory with flavors and we insert a mesonic operator, which
contains a quark and an antiquark field. Flavors correspond to adding D-branes in the bulk
\refs{\ansar,\karch}.
 The mesonic operator corresponds to an open string mode on the D-brane that is extended over
 $AdS_5$.
For example, we could consider the insertion of a flavor symmetry current which couples to
a $q, ~\bar q$ pair. This is analogous to the electromagnetic current in QCD.
Amplitudes involving quarks have been considered at strong coupling in \mcgreevy .
Once
 we IR regularize, the quarks correspond to open strings that
are attached to the D-brane and sit at $z\sim \infty$ or $r\sim
0$. The discussion is very similar to the one for closed strings.
One difference is that now we do not require the configuration to
be periodic. However, since we obey Neumann boundary conditions on
the boundary of the open string, which translate to Dirichlet
boundary conditions in the T-dual variables, we find that the
solution can be extended outside the strip into a periodic
function with a period which is twice the original width of the
strip, see figure 4 .

\ifig\fqq{Once we extend the solution outside the strip as shown in
the figure, it reduces to the zig-zag solution, with twice the period.
 } {\epsfxsize2.5in\epsfbox{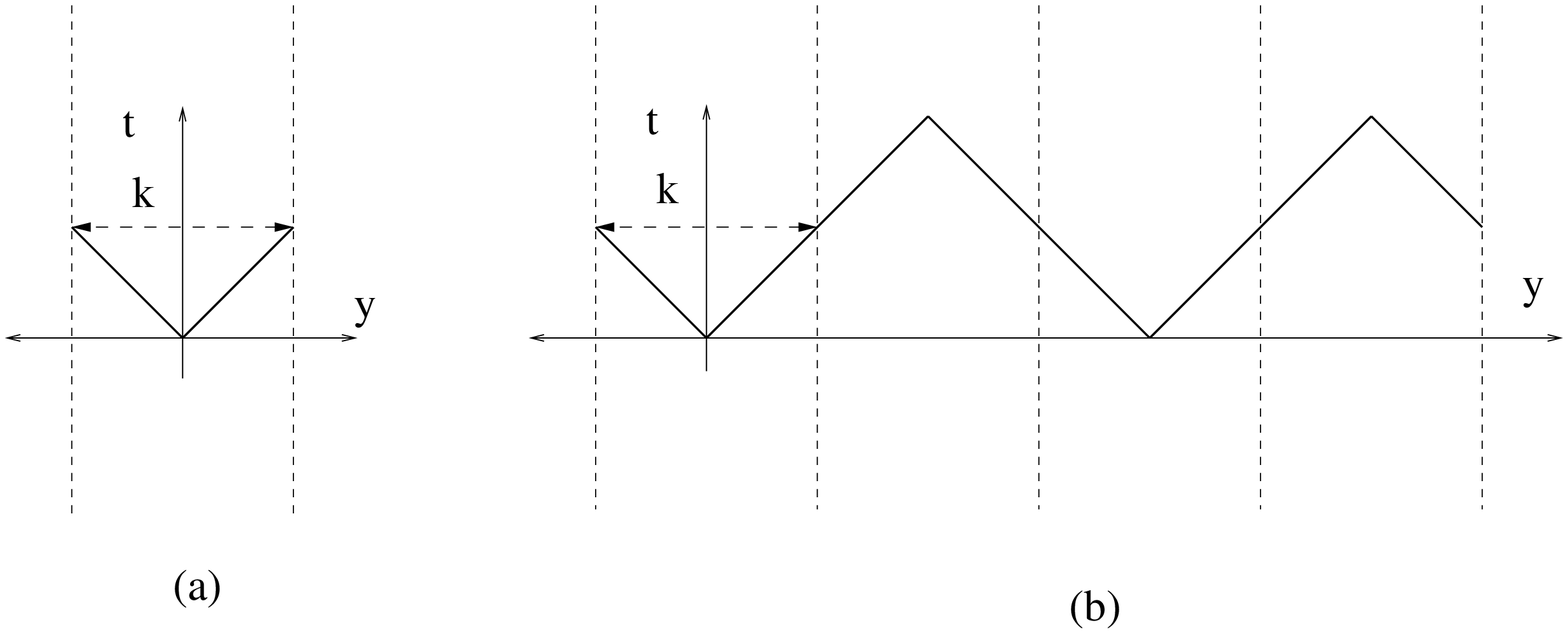}}

Thus if we consider a configuration with momenta as in \momenta ,
the solution is simply given by \eqn\newsol{ t = 2 \,  \kappa \,
\hat t \left( { r \over 2 \kappa}, { y \over 2 \kappa} \right) }
where $\hat t$ is the rescaled solution, with period one, $\hat t
( \hat r , \hat y +1) = \hat t(\hat r, \hat y) $. The action is
simply half of the action in \funfom\ but with the replacement
$\kappa \to  2 \kappa$. After we reexpress it in terms of $\kappa$
again we get \eqn\actionqbarq{ i S = - { \sqrt{\lambda} \over 2
\pi } { \mu^\epsilon \over ( 2 \pi \kappa )^\epsilon } \left[   {
1 \over \epsilon^2 } + { 1- 3 \log 2   \over  2  \epsilon}    +
\left\{ { C \over 2 } - { \log 2 \over 2} + (\log 2 )^2 \right\}
\right] } Thus we see that function $g(\lambda) $ which determines
the subleading IR divergencies is different for a gluon than a
quark. Namely, we have \eqn\ggluonquark{ g_{gluon}(\lambda ) = {
\sqrt{\lambda} \over 2 \pi } (1 - \log 2) ~,~~~~~~~ g_{quark}
(\lambda ) ={ \sqrt{\lambda} \over 2 \pi } (1 - 3 \log 2) } where
$g_{gluon}$ as computed in \fajm . In the case that we have a cusp
that joins a quark and a gluon we expect to have the average of
the above two formulas.

We can similarly consider asymptotic states corresponding to a quark and an antiquark plus
extra gluons,
 $q \bar q + n g$. In this case we simply take the configuration of momenta for all these
 particles, we flip it, and then
 take a periodic superposition as explained above. See figure 5.
The solution will be then given by half of the periodic solution
as we had above.

\ifig\fqqnn{(a) Configuration of momenta for the process $\gamma
\rightarrow q \bar q+g+g$. The dashed line with an arrow indicates
the momentum of the operator. The first and last segments are the
quark and anti-quark, the middle two segments are the two gluons.
(b) Extension of the momenta into a periodic configuration.  }
{\epsfxsize2.5in\epsfbox{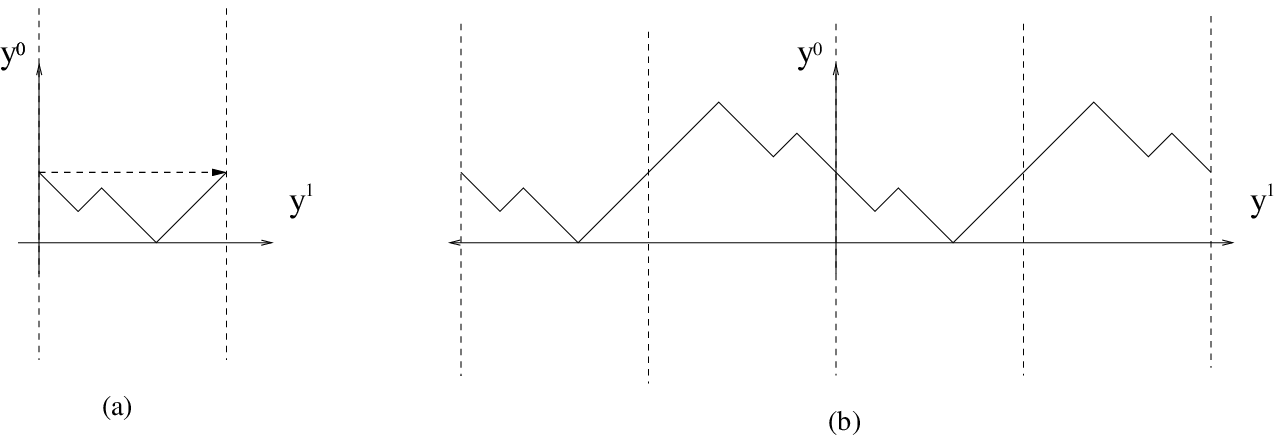}}

\newsec{ Scattering amplitudes involving a large number of gluons }

\subsec{ The Bern Dixon Smirnov ansatz}

In an interesting paper, Bern, Dixon and Smirnov \bds , proposed
the following form for the $n$ gluon planar, color ordered, MHV
scattering amplitude \eqn\formamp{
 \log \left[ {\cal A}_n \over {\cal A}_{n,Tree} \right] = {\rm Div}_n
 + { f(\lambda) \over 4} a_1(k_1,\cdots, k_n) + h(\lambda) + n k(\lambda)
 }
 where $a_1$ is the finite part of the
 one loop scattering amplitude, up to a factor of ${ \lambda \over 8 \pi^2}$.
  The divergent terms have a simple structure controlled
 by the cusp anomalous dimension $f(\lambda)$ and a second function of the coupling $g(\lambda)$.
 In dimensional regularization, $d=4 - 2 \epsilon$, we have \bds\
 \eqn\divter{\eqalign{
 {\rm Div}_n =&  - \sum_{i=1}^n  \left[ { 1 \over 8 \epsilon^2 } f^{(-2)}
 \left( \lambda \mu^{2 \epsilon}
 \over s_{i,i+1}^{\epsilon} \right) + { 1 \over 4 \epsilon} g^{(-1)}\left( \lambda \mu^{2 \epsilon}
 \over s_{i,i+1}^{\epsilon} \right) \right]
 \cr
 &~~~~~\left( \lambda { d \over d \lambda } \right)^2 f^{(-2)}(\lambda) =f(\lambda) ~,~~~~~~
 \lambda { d \over d \lambda }  g^{(-1)}(\lambda) =g(\lambda)
 }}
 where $\mu = \mu_{IR}$ is the IR regularization scale.
 $h(\lambda)$ and $k(\lambda)$ are functions of
 only $\lambda$ and will not be interesting for us. We will focus on $a_1$ which is the part
 of the amplitude that depends in a non-trivial way on the kinematic invariants of the
 process. Note that $f(\lambda)$ is a known function \bes\  which behaves as
 \refs{\KotikovER,\GubserTV}, see also \otherfun ,
 \eqn\formfas{
 f(\lambda ) = \left\{ \eqalign{
  & { \lambda \over 2 \pi^2} \left(1 - { \lambda \over 48 }  + \cdots \right)  ~,~~~~~~~~{\rm for} ~~~ \lambda \ll 1
  \cr
 &  { \sqrt{\lambda} \over \pi } + \cdots  ~,~~~~~~~~{\rm for} ~~~ \lambda \gg 1
 } \right.
 }

 The results for four and five gluons ($n=4,5$) are determined
by the momentum space (or T-dual) conformal symmetry
\refs{\wilsonplanar,\nfive}. Thus, their form follows from the
form of the leading IR divergence which is controlled by the
function $f(\lambda)$. Thus, in order to do a non-trivial test of
the BDS guess we need to perform a higher loop computation for
$n\geq 6$, where the results are not fixed by the dual conformal
symmetry. In what follows below, we will do a computation at large
$\lambda$ for large $n$ and we will find a disagreement with
\formamp .


In a recent paper Brandhuber et al \brandhuber , have shown that
the finite part of the one  loop scattering amplitude
$a_1(k_1,\cdots,k_n)$ is equal to the finite part of the one loop
expectation value of a Wilson loop consisting of $n$ lightlike
segments specified by the momenta (for $n=4$ this was shown in
\wilsonplanar ). This relation has been verified explicitly by
computing both sides and checking that they are the same. In other
words, we consider a Wilson loop specified by light-like segments
proportional to the momenta $k^\mu_i$. We consider its expectation
value to obtain \eqn\expecva{ \langle W_{\{ k_i \} } \rangle = 1
+ { \lambda \over 8 \pi^2 }\left[ {\rm Div}' + w_1(k_1,\cdots ,
k_n)  + c + n c' \right] } Then the result of \brandhuber\ is that
$w_1(k_1, \cdots , k_n) = a_1(k_1,\cdots,k_n)$. The divergent
terms arise from UV divergencies at the cusps of the Wilson lines.
They have a form similar to \divter\ but with $\mu_{IR} \to
\mu_{UV}$ and with a different function $g(\lambda)$ \wilsonplanar
\brandhuber . Thus we can say that the BDS ansatz is simply saying
that the $n$ gluon scattering amplitude is the same as the
exponentiation, up to the function $f(\lambda)$, of the one loop
expectation value of a Wilson loop. In other words, we can write
\eqn\bdsguess{ \left. { {\cal A} \over {\cal A}_{Tree} }
\right|_{BDS}=
 e^{ \rm Div}  e^{ { f (\lambda) \over 4 } w_1(k_1, \cdots , k_n)}
 e^{\tilde h(\lambda) + n \tilde k(\lambda)}
 }
where $\tilde h(\lambda)$ and $\tilde k(\lambda)$ are functions
which depend only on $\lambda$ but are independent of the shape of
the Wilson loop and will not be interesting for us.

 We now consider the strong coupling
form of the amplitude. After we go to the T-dual AdS space we find
that the computation is formally equivalent to computing the
expectation value of a Wilson loop, at least to leading order in
the $1/\sqrt{\lambda}$ expansion. On the other hand we know that
the strong coupling result for a general
  Wilson loop is not simply given by exponentiating the one loop value and multiplying
  by the function $f(\lambda)$.
In particular, let us recall the expectation value for a spacelike
rectangular Wilson loop with long length $T$ and short length $L$,
with $T/L \gg 1$. This is the Wilson loop that is useful for
computing the quark anti-quark potential. Sometimes one considers
the long side along a timelike direction. Here we consider both
sides of the rectangle along spacelike directions, so that time is
orthogonal to the loop.

\ifig\Rlines{Configuration of gluons approximating a
rectangular Wilson loop. Here $n=60$ and $T=2L$. The rectangle lies along two
spatial dimensions and the zig-zag motion is into the time direction.
 } {\epsfxsize2in\epsfbox{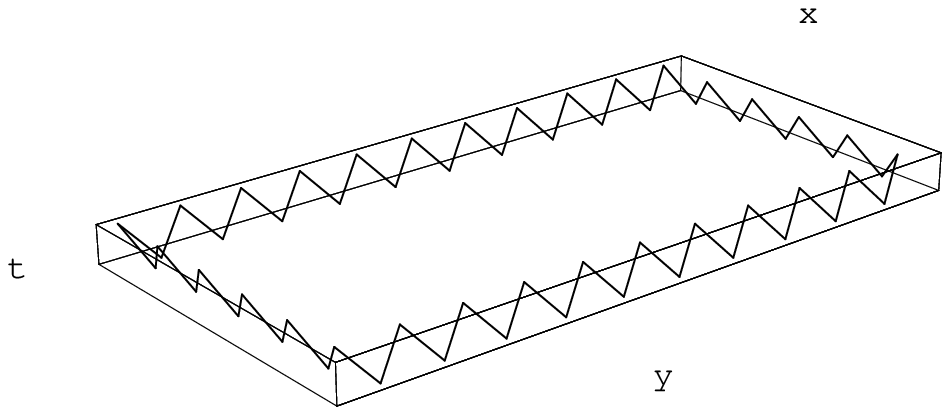}}

On the other hand, the Wilson loops that appear in amplitude computations
 consist of light-like
segments that are are not obviously related to the  rectangular spacelike
Wilson loop. However, one can approximate a spacelike Wilson loop
via a Wilson loop with lightlike segments. For instance,  one can approximate a
straight spacelike line with a zig-zag path of lightlike segments,
such as the one shown in \zigzag (b). Thus, we can approximate the rectangular Wilson
loop via the configuration of light-like lines shown in figure \Rlines .

If the zig-zag segments have spatial length $\kappa$, then we
expect that at distances longer than $\kappa$ the loop will
look like a spacelike one. In fact, in the leading  weak coupling
approximation, $w_1$,
 this is very simple to see,  since the one loop
approximation to the Wilson loop is  given by integrating the
propagator between two points along the loop. As long as the two
points are far away the zig-zag motion averages out and we recover
the same result as for a spacelike loop, up to terms that can be
viewed as the local divergencies that appear when we consider a
spacelike loop. Note that any two points on this zigzag loop that
are not on the same segment are spacelike separated.
In addition, note that the spacelike Wilson loop that we get in the limit is the one with
no scalar couplings, namely the ordinary Wilson loop operator $W = Tr[ P e^{i \oint A} ]$, we will
return to this point later.
The weak coupling result for the rectangular loop, for $T/L \gg 1$, is
\eqn\weakcoup{
\log \langle W \rangle =
 { \lambda \over 8 \pi } {T \over L}\left( 1 - { \lambda \over 4 \pi^2 } + \cdots
\right) ~,~~~~~~~~~~~~\lambda \ll 1
}
where the two loops term is computed in appendix B. Note that the coefficient of
the one loop term is half of the value for the locally
BPS Wilson loop that contains a coupling to a scalar\foot{The two loop term was
computed for the locally BPS Wilson loop in \zarembowillo , see also \pineda . }.

We now consider the configuration at strong coupling.
  Near the zig-zag path the solution will look like the
solution discussed in section 2, see \approxzigzag  . As we move
along the radial $AdS_5$ direction into the bulk the surface
approaches very quickly the solution we would have for a straight
spacelike Wilson loop, see the discussion around \tsmall . More
quantitatively, if we introduce a zig-zag path with segments of
spatial length $\kappa/2$ the classical area behaves as
 \eqn\lightlik{
 {\rm
 Area} = { \ell \over \kappa } { \mu^\epsilon \over
( 2 \pi \kappa)^\epsilon } \left( { 2 \over \epsilon^2 } + { (1-\log 2) \over \epsilon }
+ C \right) + A_{ren} + \cdots
 }
 where $\ell$ is the length of the spacelike loop that we are approximating.
 The constant $C$ is the same as in \funfom , but its value is not important for us.
 Finally,  $A_{ren}$ is the value we would
have for a spacelike Wilson loop.
 Namely, for a spacelike Wilson
loop we have \eqn\spacel{ {\rm Area} =   { \ell \over a } + A_{ren}
} where we have set a cutoff at $z=a$ and taken the limit $a\to 0$
to define $A_{ren}$.  We see that after we subtract the terms going like $1/\epsilon^2$ and
$1/\epsilon$ which are the
  IR divergencies in the amplitude in \lightlik\ we
are left with a term that is finite in the $\kappa \to 0$ limit, $A_{ren}$,
and a term that diverges, proportional to $C$. This term, which goes like ${ \ell \over \kappa} C$,
 can be interpreted as the local
divergence that appears when we  define a spacelike Wilson loop, see
\spacel . This divergence is non-universal and could depend on how we define the spacelike
Wilson loop. This divergent term   is proportional
to the perimeter length of the loop and does not depend on the
macroscopic shape of the loop. In the particular case of the
rectangular loop, the divergent term does not depend on $T/L$. Therefore we can subtract it
to obtain a finite value, $A_{ren}$, which should coincide with the finite value of
the spatial Wilson loop expectation value. More precisely, we
 get $\langle W \rangle \sim e^{ - { \sqrt{\lambda }
\over 2 \pi } A_{ren} }$ to leading order in $1/\sqrt{\lambda}$.

Since the light like lines had no coupling to the scalars, the Wilson loop that
we approximate in this way is the Wilson loop with no scalar coupling. However, as we
show in section 4, to leading
order in the $1/\sqrt{\lambda} $ expansion, the two results are the same.
 Both are given in terms
of the same minimal surface in $AdS_5$.
 Then we find that the result for the rectangular loop in the regime $T/L \gg 1$  is given by
 \refs{\jmwilson,\wilsonrey},
 \eqn\wilsonstrong{
  \log \langle W \rangle =
  \sqrt{\lambda} { 4 \pi^2  \over \Gamma({1\over
4})^4} {T \over L} ~,~~~~~~\lambda \gg 1
}

The final conclusion is that for a particular configuration of gluons, where $n\to \infty$ we
can approximate
 both the BDS guess and the large $\lambda$ results by considering a spacelike loop.
In this limit, if the BDS result were true, we would obtain a different numerical coefficient
for the result of a Wilson loop expectation value. Instead of the second line in
\wilsonstrong\ the BDS guess would produce $\log \langle W \rangle
= { \sqrt{\lambda} \over 4} {T\over L}$, were we used
the known strong coupling form for the function $f(\lambda)$ \formfas . We should also
mention that the functions $h(\lambda)$ and $k(\lambda)$ that appear in \formamp\ cannot
fix the disagreement since they are independent of the shape of the Wilson loop.

The reader might worry about the following. The strong coupling prescription for the
Wilson loops that we are using is simply the leading order term in the $\sqrt{\lambda}$
expansion. At the next order we can have further fields that propagate on the worldsheet that
encode the polarization states of the asymptotic states. Since we are taking the $n\to \infty$
limit, one might worry that such terms might become important. In particular, we have not used
the fact that we are considering an MHV amplitude. In order to avoid this problem
we take  the
large $\lambda$ limit and then the large $n$ limit, so that $\sqrt{\lambda } \gg n  \to \infty$.
With this particular order of limits we do not have to worry about such subleading corrections.
We can take the same limit of the BDS guess \bds , which was supposed to be valid for all values
of $\lambda$ and $n$.

Of course, it would be nice to find out exactly at what order in $\lambda$ the first deviation appears.
For this purpose we should consider an amplitude for $n\geq 6 $ external particles.

\subsec{Scattering amplitudes and Wilson loops }

Recently, another conjecture has been entertained
  \refs{\wilsonplanar,\brandhuber,\nfive}, which agrees with what
  we know presently.
   The conjecture is that the finite part of the planar color ordered
    MHV amplitude is given
by the finite part of
the Wilson loop expectation value in the full theory
\eqn\wilsoexp{ { { \cal A}_n \over {\cal A}_{n,Tree} } = e^{ Div}
\langle W_{k_i} \rangle e^{- Div'}  e^{\hat h(\lambda) + n \hat k(\lambda) }
}
Where Div corresponds to the IR divergencies of the amplitude,
while ${\rm  Div}'$ is subtracting the UV divergencies of the Wilson
loop. The functions $\hat h(\lambda)$ and $\hat k(\lambda)$ are independent of
the shape of the Wilson loop.  The statement is only about the momentum dependent
finite
remainder of the amplitude.
Of course, this does not give us an explicit function since we do
not know how to independently compute the expectation value of the
Wilson loop.

 {}From the string theory point of view this conjecture implies that
computing the Wilson loop expectation value in the T-dual $AdS_5$
space gives the same answer as computing it in the original
$AdS_5$ space. At leading order in the $1/\sqrt{\lambda}$
expansion, this is obvious since all we do in both cases is to
compute the area of the minimal surface in $AdS_5$. On the other
hand, to higher orders in the $1/\sqrt{\lambda}$ expansion we
would start noticing that  the RR fields and the dilaton are
different in both cases. Thus, the conjecture amounts to the
statement that we can perform a redefinition of the worldsheet
fermions in such a way that the  two backgrounds lead to the same
result. It would be interesting to check  explicitly whether this is
true or not.
Note that if the ''dual'' momentum space conformal
 symmetry \refs{\firstconformal} is an exact symmetry of the
theory, then the agreement between Wilson loops and amplitudes for
$n=4,5$ would be obvious since in both cases the answer is
determined by the symmetries \refs{\wilsonplanar,\nfive}.

As another remark in the same direction, we note that the two loop
result for the rectangular Wilson loop at weak coupling, \weakcoup
, is not given by the two loop correction to the function
$f(\lambda)$ \formfas . This implies that the two loop expression
for the Wilson loop is not given by the exponential of the one
loop result (up to the function $f(\lambda)$) for $n$ large. In
turn, this implies that either the relation between amplitudes and
Wilson loops is not correct at two loops or that the BDS guess for
the amplitude is not correct at two loops\foot{ It could well be
that the BDS guess for the amplitude is correct at two loops for
all $n$, but that the relation between Wilson loops and the
amplitudes is incorrect at two loops.}. In any case, we see that
we would learn something from an explicit two loop computation of
the amplitude for $n\geq 6$.

\newsec{String dual of the ordinary Wilson loop}

In this section we discuss our proposal for the string dual of the
ordinary Wilson loop
operator
 \eqn\ymwl{
  W ({\cal C}) = { 1 \over N} Tr[P e^{i\oint_{\cal C} \dot{x}^\mu
 A_\mu} ]
 }
Note that in  ${\cal N}=4$ super Yang Mills one often considers the
locally 1/2 BPS Wilson loop which contains a coupling to scalar
fields
 \eqn\ymwlscal{
 W_{BPS}({\cal C})= { 1 \over N} Tr[P e^{\oint_{\cal C} i \dot{x}^\mu
 A_\mu + |\dot{x}|\theta^i \Phi^i }]
 }
 where
 $\phi^i$ are the six scalar fields of ${\cal N}=4$ SYM and
$\theta^i$ a unit vector in $R^6$. At strong coupling we know that
the description of the Wilson loop \ymwlscal\ is given by a string worldsheet
that ends on the boundary of $AdS_5$ on the corresponding contour with
Dirichlet boundary conditions on the sphere which force the loop to sit at the
point $\theta^i$ \refs{\jmwilson,\wilsonrey,\grossetal}.

The above discussion suggests that we can approximate
a
Wilson loop with no coupling to the scalars, such as \ymwl\ by a ziz-zag Wilson
loop. Note that the light-like Wilson loops do not have couplings to the scalars.

 This leads to a natural conjecture
  for the string dual of the ordinary Wilson loop.  Namely, the ordinary Wilson loop
  operator, \ymwl , is described by a string worldsheet that ends on the boundary
  of $AdS$ on the loop ${\cal C}$
   and has Neumann boundary conditions on the five-sphere.
   The idea that this loop is somehow  related to Neumann conditions also
   appeared in \grossetal , but an explicit  prescription was not
   stated.

We make a simple consistency check of this relation.
Let us consider the gauge theory on $S^3 \times R$ and consider a
Wilson line for a quark sitting at a point on the $S^3$ and an antiquark at the opposite point.
At strong coupling such loop corresponds to a worldsheet which is $AdS_2$ and is embedded in $AdS_5 $.
One can check that it costs finite energy to move the position of the string on the $S^5$.
It is consistent with the equations of motion of
the string to set Neumann boundary conditions. In fact, if we consider small fluctuations
on the $S^5$ we see that these are described, to leading order,
 by massless fields that live on $AdS_2$.
In other words, we have massless fields living on a strip of length $\pi$.
Thus, the dimension of operators with spin $J$ in $SO(6)$ is given by
\eqn\inteq{
\Delta = { J (J+4) \over \sqrt{\lambda}  } + o({ 1 \over \lambda} )
}
which arises simply by considering  the quantum mechanics
of the center of mass motion, which is
the lowest mode on the strip, when we have Neumann boundary conditions.

We see that we get finite energies for states whose endpoints are moving on the
$S^5$. In fact, if we consider the simplest case, with $J=1$, we see that the
energy \inteq\ can be interpreted as the anomalous dimension, at strong coupling,
 for the insertion of
an operator $\phi^i$ along the loop. Namely, we consider an operator of the form
 \eqn\formop{
 P e^{ i\int_{-\infty}^0 A } \phi^i(0) P e^{i\int_{0}^\infty A }
}
 If we consider a straight contour which preserves
an $SL(2)$ subgroup of the conformal group which includes the dilatation operator, then
we see that we can assign well defined scaling dimensions to such operators along a Wilson
loop. These scaling dimensions then determine the correlation functions for two insertions. In general, when we have two insertions of such operators
we have
\eqn\correl{
\langle P[ {\cal O}(0) {\cal O}(a) e^{i\int A } ] \rangle =
{ C \over a^{2 \Delta} } \langle Pe^{i\int A} \rangle
}
Using standard perturbation theory we can compute the leading correction to the anomalous
dimension for the $\phi^i$ insertion and we obtain (see appendix B for more details)
\eqn\anomdim{
\Delta = 1 - {  \lambda \over 8 \pi^2 } + o(\lambda^2)
}
Thus we see that already to leading order in the weak coupling expansion the anomalous
dimension tends to go down, and we observe that at strong coupling this
anomalous dimension is close to zero since it goes like $1/\sqrt{\lambda}$.

This should be contrasted with the results that are obtained for
the 1/2 BPS Wilson loop. Let us say that we consider a loop which
couples to the operator $\phi^6$ (the operator \ymwlscal\ with $\theta^6=1$).
Then the insertion of the
operators $\phi^1, \cdots, \phi^5$ corresponds to BPS operators of
dimension $\Delta =1$, see the discussion in \drukkerwil .
 On the other hand the dimension of
insertions of $\phi^6$ do not correspond to any obvious light
field on the string worldsheet\foot{We thank L. Yaffe for asking us this question.}.
Thus it is natural to think that at
strong coupling its dimension will be of the order of the mass of
the typical string state which goes like $\lambda^{1/4}$ as in
\gkpfirst . Indeed, when we compute the weak coupling anomalous dimension of
this operator we find that it is not protected with the following anomalous
dimension
\eqn\weakc{
\Delta = 1 + {\lambda \over 4 \pi^2} + o( \lambda^2 )
}
So we see that quantum corrections increase the anomalous dimension. Presumably this
continues to increase so that at strong coupling it has a value larger than $\lambda^{1/4}$.

Notice that, to leading order,  the strong coupling result for the
 Wilson loop is the same as for the
 locally supersymmetric Wilson
  loop with a constant
 $\theta^i$.
  However, we will find differences at the next order in the
 $1/\sqrt{\lambda}$ expansion since in the case of the ordinary loop we would have to integrate
 over the point on the sphere where the string is sitting. More precisely, at
 the one loop level in the $\alpha'$ expansion we find that the
 determinants for quadratic fluctuations are different in the two cases \DrukkerEP .

On the other hand, the weak coupling result is rather different for the two loops. For example
the leading term in the quark anti-quark potential for the ordinary Wilson loop is half the
value we have for the locally half BPS loop since we only have the gauge boson exchange, rather
than the boson exchange plus the scalar exchange.
In fact, we also computed the two loop correction for the quark anti-quark potential and we
found the result shown in \weakcoup . The two loop correction for the quark anti-quark potential
coming from the half BPS Wilson loop was computed in \refs{\zarembowillo,\pineda}, where
a logarithmic term of the form $\log \lambda$ was found at this order.

\newsec{Conclusions}

In this article we have given a prescription for computing
processes involving local operators and single gluon asymptotic states at
strong coupling in ${\cal N}=4$ super Yang Mills.
As in the case where we only have gluons, the problem reduces to finding a classical
surface in a T-dual $AdS_5$ space. The surface ends on a contour on the boundary of this T-dual AdS space which is specified by the
momenta of the gluons. In addition the surface also goes to $ r \to \infty$ which corresponds
to the boundary of the original $AdS_5$ space. This boundary condition characterizes the operator
in question.


We have then proceeded to test the BDS \bds\ ansatz for a particular configuration involving a large number of
gluons $n \to \infty$. We found that it did not survive the test.
We took the $n\to \infty $ limit to simplify the computations. However, we think that it is
likely that already for $n=6$ the strong coupling answer will not have the BDS form \bds .

On the other
hand, recently \refs{\wilsonplanar,\brandhuber,\nfive} an interesting
equivalence between two seemingly different
observables was considered: scattering
amplitudes and Wilson loops.  Our results are consistent with such relation.
However, our weak coupling computation for the rectangular Wilson
loop suggests that something new is happening at two loops in the
weak coupling expansion. We found that the Wilson loop vacuum
expectation value
 is not given by the BDS ansatz. The reason could be that the Wilson loop is not equal
to the amplitude or that the BDS ansatz is not the correct value
of the amplitude at two loops. So,   we would learn something new
by computing the amplitude at two loops for $n\geq 6$.

We have also considered some features of the ordinary Wilson loop in ${\cal N}=4$ super Yang
Mills. This is the Wilson loop with no scalar couplings. We described the string prescription
that corresponds to this Wilson loop. We have also computed the anomalous dimension for the
insertion of a scalar field at a point on this Wilson loop, both at strong and weak coupling.
We have also computed the static potential for the ordinary Wilson loop to two loops.
This result does not satisfy the exponentiation properties suggested by the BDS conjecture.
 As mentioned above, this suggest that either the BDS conjecture or the relation between Wilson loops and scattering amplitudes fails at two loops for a large enough number of gluons.

{\bf Acknowledgments}

  We would like to thank
Z. Bern, L. Dixon,  G. Korchemsky, H. Liu, J. McGreevy,
  G. Sterman,  E. Sokatchev, M. Spradlin, A. Volovich and
L. Yaffe
 for discussions.

This work   was  supported in part by U.S.~Department of Energy
grant \#DE-FG02-90ER40542. The work of L.F.A was supported by VENI
grant 680-47-113.

\appendix{A}{ Some remarks on the worldsheet solution for five gluons.}

At strong coupling, the amplitude is computed by consider a Wilson
loop with five light-like segments that end on five points. The
data specifying the configuration corresponds to the five points
${x_1, \cdots , x_5}$ which are such that $( x_i - x_{i+1})^2 =0$.
One can easily check that there are no invariants that can be made
out of these five points. In particular, by a conformal
transformation we can always map points $x_1 \to 0$ and $x_2 \to (
1,1, \vec 0)$. Then the other three points are on the light-like
boundary of Minkowski space. Another way to see this is that we
have $4 \times 5 - 5 = 15$ parameters and this is the same as the
dimension of the conformal group. This argument also shows that
when we go to $n=6$ we expect that conformal symmetry will {\it
not} determine the answer.

As in the case of $n=4$   \wilsonplanar\ one can show that the
$n=5$ Wilson loop is completely determined by conformal
symmetry\foot{We thank M. Spradlin and A. Volovich for a
conversation on these issues.}  \nfive . Of course, the fact
that we get a non-trivial function comes from the fact that we
have divergencies at the cusps, otherwise we would get a result
that is independent of the momenta. Thus we do not need to find
the explicit surface to know that the strong coupling answer will
agree with the BDS conjecture \bds , for $n=5$.
 Nevertheless let us make a
few remarks on finding the surface explicitly. The surface is
simplest to understand if we have a spacelike momentum transfer at
all the cusps. Other cases could  be found by an appropriate analytic
continuation.

\ifig\threelines{Wilson loop in $R^{2,2}$. We display the contours
in the $(x_1,x_2)$, $(t_1,t_2)$ and $(x_1^+,x_1^-)$ plane
respectively of a simple configuration corresponding to the
scattering of 5 gluons. The remaining three points are located on
the light-like boundary of $R^{2,2}$.}
{\epsfxsize3.5in\epsfbox{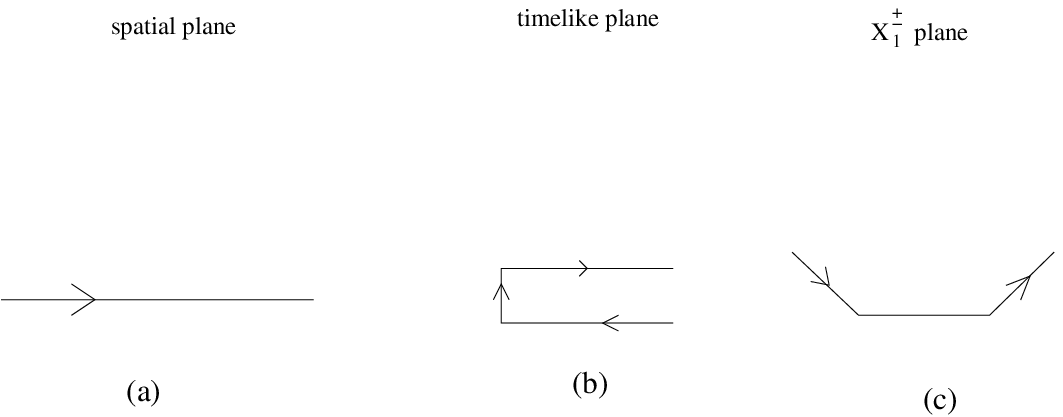}}

In 1+3 signature it is not possible to have a real configuration
of five light-like momenta so that the sum of two consecutive ones
is always spacelike. On the other hand this is possible for real
momenta in 2+2 signature. This corresponds to complex momenta in
1+2 signature.  Thus it is convenient to look for the surface
using 2+2 signature, where there is a completely real solution.
 Since all configurations are related by
conformal symmetry, we can attempt to find the simplest one. A
simple configuration is the one shown in \threelines . It consists
of a set of three lines in $R^{2,2}$, the other two lines live on
the light-like boundary of $R^{2,2}$.\foot{ One can relate
$R^{2,2}$ to a patch of $S^2 \times S^2$ in the same way that
$R^{1,3}$ is embedded into a patch of $R \times S^3$. This is done
as follows. We write $ds^2 = - dr_1^2 - r_1^2 d \varphi_1^2 + dr_2^2
+ r_2^2 d \varphi_2^2 = { 1 \over 4 \cos^2 v^+ \cos^2 v^-} ( -
d\theta_1^2 - \sin^2 \theta_1 d\varphi_1^2 + d\theta_2^2 + \sin^2
\theta_2 d\varphi_2^2)$, where $r_1 \pm r_2 = \tan v^\pm$ and
$v^\pm = (\theta_1 \pm \theta_2 )/2 $.
 } The three
lines are parametrized by $(t_1,t_2,x_1,x_2)$ given by
\eqn\paramlin{\eqalign{
 & ( \lambda, -1, - \lambda , 0) ~,~~~~~\lambda \geq 1
 \cr
 & ( 1, \lambda , \lambda , 0 ) ~,~~~~~~~~ -1 \leq \lambda \leq 1
 \cr
 & ( \lambda , 1, \lambda , 0 )~,~~~~~\lambda \geq 1
 }}
We can see that $x_2=0$ everywhere and we can view the solution as $t_2(x_1^+,x_1^-)$ where
$x_1^\pm = t_1 \pm x_1$.
The behavior near each of the cusps is given by the single cusp solution in
\kruczenski .
In order to gain some intuition of this solution, we can study the
equations of motion in the regime of large $x_1^\pm$. In this
regime $t_2$ is small and we can linearize the equation
\eqn\eomlargex{\left(\partial_\tau^2+\partial_\chi^2-2\partial_\tau
\right)t_2(\chi,\tau)=0} where $x^{\pm}=e^{v^{\pm}}$ and
$v^\pm=\tau \pm \chi$. It seems natural to assume that for large
$\tau$ the function behaves as $t_2=t_2(\chi/\sqrt{\tau})$  and
assume that $u=\chi/\sqrt{\tau}$ is kept fixed as $\tau$ becomes
large. Then we find the following form for the solution
\eqn\sollargex{u \partial_u t_2(u)+\partial_u^2
t_2(u)=0,~~~~\rightarrow~~~~t_2(u) \approx {\rm Erf}(u/\sqrt{2}) } The
presence of the error function suggest that the full solution may
not have a simple algebraic expression.

\appendix{B}{Some perturbative computations involving Wilson loops}

\subsec{One loop anomalous dimensions of scalar insertions}

We normalize the action so that there is an overall $1/g^2$ term.
Then the propagator of the scalar fields has the form \eqn\propag{
\langle \phi_{~i}^{j}(x) \phi_{~k}^{l}(0) \rangle = { g^2 \over 8
\pi^2 } { \delta_{i}^l \delta_k^j \over x^2 } } and  $\lambda = g^2 N$.
\ifig\diagrams{In (a) we see a local operator insertion along a
Wilson loop. In (b) we see two operators insertions leading to a
two point function. In (c) we see the tree level contribution to
the two point function. In (d),(e) we see diagrams that determine
the contribution of the anomalous dimension of a $\phi^1$
insertion along an ordinary Wilson loop. These are diagrams that
end up contributing (with a minus sign) to the BPS wilson loop.
Thus, they contribute with a plus sign to the ordinary loop, as
explained above.
  In (f),(g),(h) we see various
diagrams that contribute to the two point function of the operator $\phi^6$ which also
appears
in the expression of the BPS  Wilson loop. These diagrams contribute with minus signs because
they represent the contributions that we oversubtract when we subtract the exponentiation of
the diagrams in
(i),(j) which would lead to a one point function.  } {\epsfxsize3.5in\epsfbox{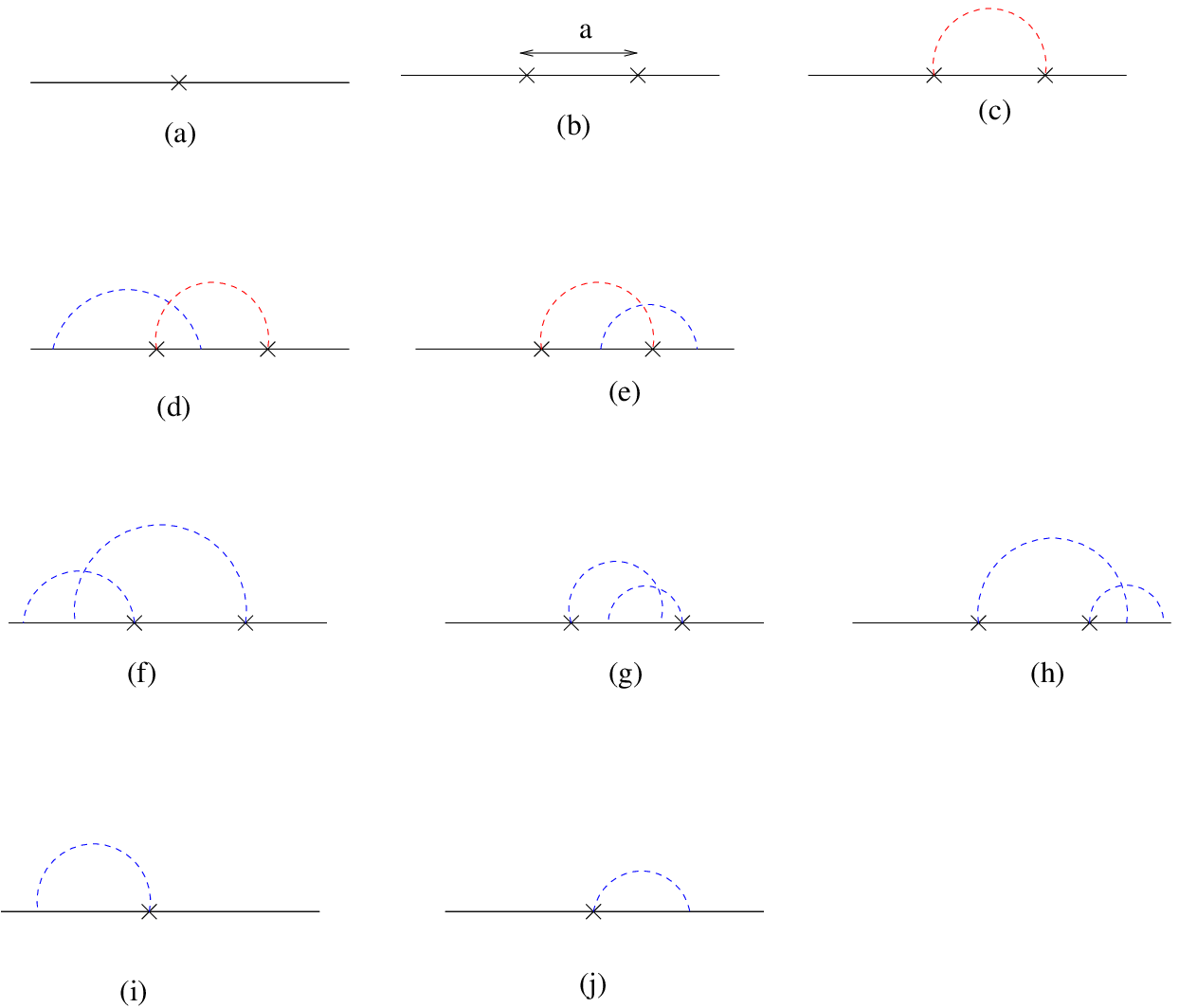}}
Let us first consider the one loop computation of anomalous dimensions for local operators
inserted on the loop. We compute them by considering the correlation function
\eqn\correlap{
\langle P[ {\cal O}(0) {\cal O}(a) e^{i\int A } ] \rangle =
{ C \over a^{2 \Delta} } \langle Pe^{i\int A} \rangle
}
where $C$ is a constant and $P$ denotes ordering along the path.
Let us now compute the anomalous dimension for an insertion of an $SO(6)$ scalar $\phi^i$
along the ordinary Wilson loop \ymwl .
At lowest order we simply have the contribution in \diagrams (c). At the next order in
$\lambda$ we get several diagrams. In order to simplify the computation we note that
very similar diagrams appear when one considers the correction to the anomalous dimension
for an insertion of $\phi^1$ into a BPS Wilson loop which involves $\phi^6$, such at
\ymwlscal\ with $\theta^6=1$. For the BPS loop the $\phi^1$ insertion is BPS and its anomalous dimension is  protected. Thus, the only diagrams that
can contribute to the anomalous dimension in our case come from diagrams which appear for
the BPS loop but do not appear for the ordinary loop  or viceversa. The only such diagrams are
diagrams involving contractions of the scalar field $\phi^6$ which appears in the exponent of the
BPS Wilson loop. At this order we do not have contractions between the inserted field, $\phi^1$, and
the field appearing in the exponent, $\phi^6$.
 However, the inserted field affects the contractions
of the scalar that appears in the loop because of the planarity restriction. Note that
those contractions also appear in the right hand side of \correlap\ when we compute the
expectation value of the loop without an insertion. Thus we need to
 consider diagrams which represent contributions that are in the right hand of the BPS loop
side but are not in the left hand side of the BPS loop, due to planarity.
Such diagrams are shown in
figure \diagrams (d),(e). These diagrams contribute to a term of the form
\eqn\teermf{
\langle P  \phi^i(0) \phi^i(a) e^{i \int A} \rangle =  { C \over a^2 } \left(
1 + { \lambda^2 \over 4 \pi^2 } \log(a/\epsilon) \right) \langle e^{i\int A }\rangle
}
This implies that the anomalous dimension is \anomdim .

In a similar way we can compute the anomalous dimension for the insertion of a $\phi^6$
field on a half BPS Wilson loop which couples to $\phi^6$ (i.e. \ymwlscal\ with $\theta^6=1$).
  This operator will have
a non-zero expectation value $\langle \phi \rangle $ which will come from contractions between
the insertion and the fields along the loop, see \diagrams (i),(j).
These should be subtracted to define a good
conformal operator, whose expectation value is zero.
So from now on we consider the operator defined with these subtractions.
 As before, we only have to consider the diagrams which appear when we insert a $\phi^6$ but
 do not appear when we insert a $\phi^1$ or viceversa.
   Such diagrams are the ones where the insertion is
contracted with the fields appearing in the loop.
There are many diagrams of this kind. However, all the planar diagrams are subtracted when
we subtract the vacuum expectation value of $\phi^1$ to define the operator. In fact, this
subtraction is subtracting diagrams which do not appear when we consider the two point
function without any subtraction. These extra subtractions are the only contribution and
are displayed in
 \diagrams (f),(g),(h). They thus contribute with a minus sign.
Computing explicitly such diagrams one focuses on the term going like $\log a$
from which we can extract the correction to the anomalous dimension to obtain \weakc .

 \subsec{Two loop computation of the quark antiquark potential}

In this appendix we give some details on the computation of the quark anti-quark potential
for the case that we have a quark that couples only to the gauge field, as in \ymwl .
A similar computation for the case that we also have a scalar coupling was done in
\refs{\zarembowillo,\pineda}. We perform the computation in Euclidean space.

\ifig\twoloop{All lines, except the vertical ones,
 are gluon propagators.  (a) One loop vacuum polarization correction to the gluon propagator.
(b) Non iterative ladder. This diagram represents the failure the exponentiation of the one
loop result, so it should be subtracted from the log of the Wilson loop expectation value.
 (c) Similar  non-iterative subtraction. The
factor of two because it appears on both lines. Note that the lines that cross in (b) and (c)
do not denote interaction vertices.
(d) Diagrams with an interaction  vertex that does not contribute.
} {\epsfxsize1.5in\epsfbox{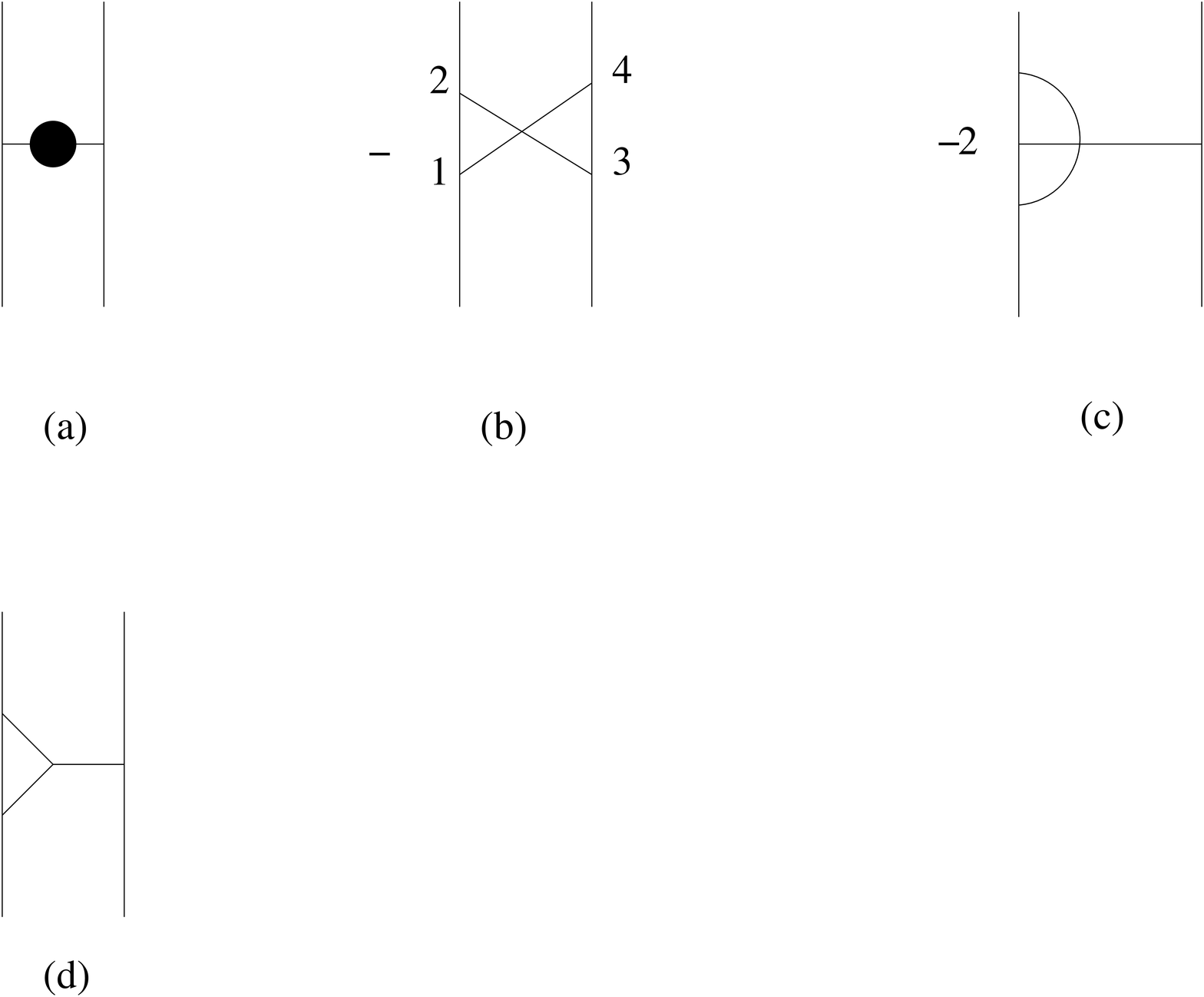}}

 The diagrams to be computed as are in figure \twoloop . The diagram in
 \twoloop (d) does not contribute in Feynman gauge because all the indices of the gluons along the loop
 are all forced to zero 0, and there is no vertex in the Yang Mills lagrangian involving three
 gauge fields with the same value of the index.
The diagrams in \twoloop (b),(c) are diagrams which do not actually appear due to the planarity
restriction, but are diagrams that need to be subtracted explicitly. In other words, it is
convenient to add and subtract the diagrams in \twoloop (b),(c). These diagrams added to the
rest of the planar diagrams give the exponentiation of the one loop result.

Diagrams \twoloop (a) and \twoloop (c) have UV divergencies that cancel. Diagrams \twoloop (b)
and \twoloop (c) have IR divergencies which also cancel each other. The diagrams \twoloop (c)
contains a single power of the distance and it is thus zero in dimensional regularization.
Thus we only need to consider explicitly the diagrams (a) and (c).
The one loop gluon propagator has the form
\eqn\gluonp{
{ g^2 \over 2 } { 1 \over p^2 } \left ( 1 - 4 \lambda { \Gamma(\epsilon) \Gamma(2-\epsilon)
\Gamma(1-\epsilon) \over ( 4 \pi)^{ 2 - \epsilon} \Gamma(3 -\epsilon)} { 1 \over p^{ 2 \epsilon} }
\right)
}
where we wend to $D = 4 - 2 \epsilon $ dimensions. We will find it convenient to use
the following formula
\eqn\foutr{
I_{\alpha, D} = \int { d^{D-1} p \over (2 \pi )^{D-1} } p^{-\alpha}
e^{i \vec p \vec y } = y^{\alpha -D +1} \pi^{ - { D-1 \over 2} } 2^{-\alpha}
{ \Gamma[ { D-\alpha -1 \over 2 } ] \over \Gamma[{\alpha \over 2 } ] }
}
We then find that the one loop correction to the gluon propagator in \gluonp\ gives us
a result of the form
\eqn\adiagre{
(a) =  - T { \lambda^2 \over 2 }{ \Gamma( \epsilon) \Gamma( 2-\epsilon) \Gamma( 1 -  \epsilon) \over
(4 \pi)^{2 - \epsilon} \Gamma(3 - 2 \epsilon)} 4 I_{2 + 2 \epsilon , D} ~,~~~~~D=4 - 2 \epsilon
}
were $T$ is the length of the dimension along the loop.

The structure of the (b) diagram is
\eqn\strucdi{
(b) = T \int_0^\infty dt_2 \int_{-\infty}^\infty dt_3 \int_{t_3}^\infty dt_4
G(t^2_4 +y^2)G((t_2-t_3)^2 + y^2 )
}
where we have set $t_1=0$ using translation invariance. We also are using the expression for
the propagator in position space $G$. We can go to Fourier space by using the following
expression for the step function $\Theta(t)$
\eqn\canwri{
\Theta(t) = { 1 \over i } \int_{-\infty}^\infty { d k \over 2 \pi } { 1 \over k - i \eta } e^{ i k t }
}
where $\eta$ is small. We can then use this integral to enforce the ordering we
have in the \diagrams (b). We find (even though we use the index 0, we are in Euclidean space)
\eqn\integr{\eqalign{
(b) &=  - { \lambda^2 \over 4 } T { 1 \over i^2}
\int dt_2 dt_3 dt_4 \int { d k \over 2 \pi } { d k' \over 2 \pi } { dq_0  d p_0 \over ( 2 \pi)^2}
{ d^{D-1} p \over (2 \pi)^{D-1} } {d^{D-1} q \over ( 2 \pi )^{ (D-1) } } \times
\cr
&
\times e^{ i q_0 t_4} e^{ i k t_2} e^{ i p_0 (t_3-t_2) }
e^{ i k' (t_4- t_3) } e^{ i (\vec p - \vec q ) \vec y }   { 1 \over k - i \eta } { 1 \over k' - i \eta }
{ 1 \over q_0^2 + q^2} { 1 \over p_0^2 + p^2 }
}}
The minus sign arises from the fact that we are subtracting this contribution.
We can now do the integrals over $t_2,t_3,t_4$ which give
$ (2 \pi)^3 \delta(k-p_0) \delta(p_0 -k') \delta(q_0 + k') $.
We now then need to the integral over $p_0$ which is of the form
\eqn\integrin{
\int { d p_0 \over 2 \pi } { 1 \over ( p_0 - i \eta)^2 (p_0^2 + p^2) (p_0^2 + q^2 ) }
= - { 1 \over 2 q^3 p^2 } - { 1 \over 2 q^2 p^3 } + { 1 \over 2  p^2 q^2 (p+q) }
 }

Thus the end result for the (b) type diagrams seems to be
\eqn\btyped{\eqalign{
(b) = &{ \lambda^2 T \over 4 } \left( -  I_{2,D} I_{3,D} +  \tilde I  + o(\epsilon)  \right) ~ ,
\cr & \tilde I = { 1 \over 2} \int { d^3 p \over (2 \pi)^3} \int { d^3 q \over (2 \pi)^3 } { e^{i (\vec p+ \vec q). \vec y}
  \over
q^2 p^2 (p+q) } = { 1 \over y } { \log 2 \over 2^3 \pi^3 }
}}
where we have set $D=4$ in the last term since it is not divergent.
Summing the contributions of (a) and (b) and taking $\epsilon \to 0$ we get
\eqn\twoloopcont{
\log \langle W \rangle   =  { 1 \over 8 \pi } \left( \lambda - { \lambda^2 \over 4 \pi^2 } + \cdots \right)
 { T \over y }
   }
   where the dots indicates terms of order $\lambda^3$ (or possibly terms of order $\lambda^3 \log
   \lambda$ as was found in QCD \dineetal ).

 \listrefs

\bye